\documentclass[aps,twocolumn,showpacs,showkeys,preprintnumbers,amsmath,amssymb]{revtex4}

\usepackage{graphicx}
\usepackage{dcolumn}
\usepackage{bm}

\begin{document}

\preprint{}

\title{Pressure-induced isostructural phase transition of metal-doped silicon clathrates}

\author{Toshiaki Iitaka}
\email{tiitaka@riken.jp}
\homepage{http://www.iitaka.org/}
\affiliation{
Computational Astrophysics Laboratory, \\
RIKEN (The Institute of Physical and Chemical Research) \\
2-1 Hirosawa, Wako, Saitama 351-0198, Japan}

\date{\today}

\begin{abstract}
We propose an atomistic model for the pressure-induced isostructural phase transition of metal-doped silicon clathrates, $\mathrm{Ba_{8}Si_{46}}$ and $\mathrm{K_{8}Si_{46}}$, that has been observed at 14 GPa and 23 GPa, respectively.  The model explains successfully the  equation of state, transition pressure, change of Raman spectra and dependence on the doped cations as well as the effects of substituting Si(6$c$) atoms with noble metals. 
\end{abstract}

\pacs{61.50.Ks, 61.48.+c}
\keywords{silicon clathrate, phase transition, high pressure, first principles calculation}
\maketitle


{\em Metal-doped silicon clathrates} M$_{8}$Si$_{46}$ are the crystals that have the same structure (type I, space group $Pm\overline{3}n$) as the {\em methane hydrate} $\mathrm{(CH_{4})_{8} (H_2O)_{46}}$ \cite{Sloan1998} and the {\em melanophlogite} $\mathrm{A_2 B_6 (SiO_2)_{46}}$ where A and B are gas molecules \cite{Gies1983}. In the cubic unit cell, two small cages (Si$_{20}$) are composed of Si atoms at 16$i$ and 24$k$ sites (see \cite{bilbao} for crystallographic notations); one cage is located at the corner and the other is at the body-center of the cell. The small cages are bridged by the direct covalent bonding between the Si(16$i$) atoms as well as by the covalent bonding via six additional Si atoms at the 6$c$ sites, thereby forming six large cages (Si$_{24}$) in the space between the small cages. Each small and large cage contains a metal atom M at its center, i.e., 2$a$ sites and 6$d$ sites, respectively. These metal-doped silicon clathrates are attracting much attention as nano-functional materials due to their remarkable properties in superconductivity\cite{Yamanaka2000,Kawaji1995,Tanigaki2003,Tse2005}, thermoelectricity\cite{Cohn1999,Tse2000}, magnetism\cite{Kawaguchi2000} etc. The high-pressure study of these materials are especially important because synthesis of varieties of these materials has become possible by using high pressure method \cite{Yamanaka2000,Reny2000}.

\begin{figure}
\begin{center}
\resizebox{0.5\textwidth}{!}{\includegraphics*{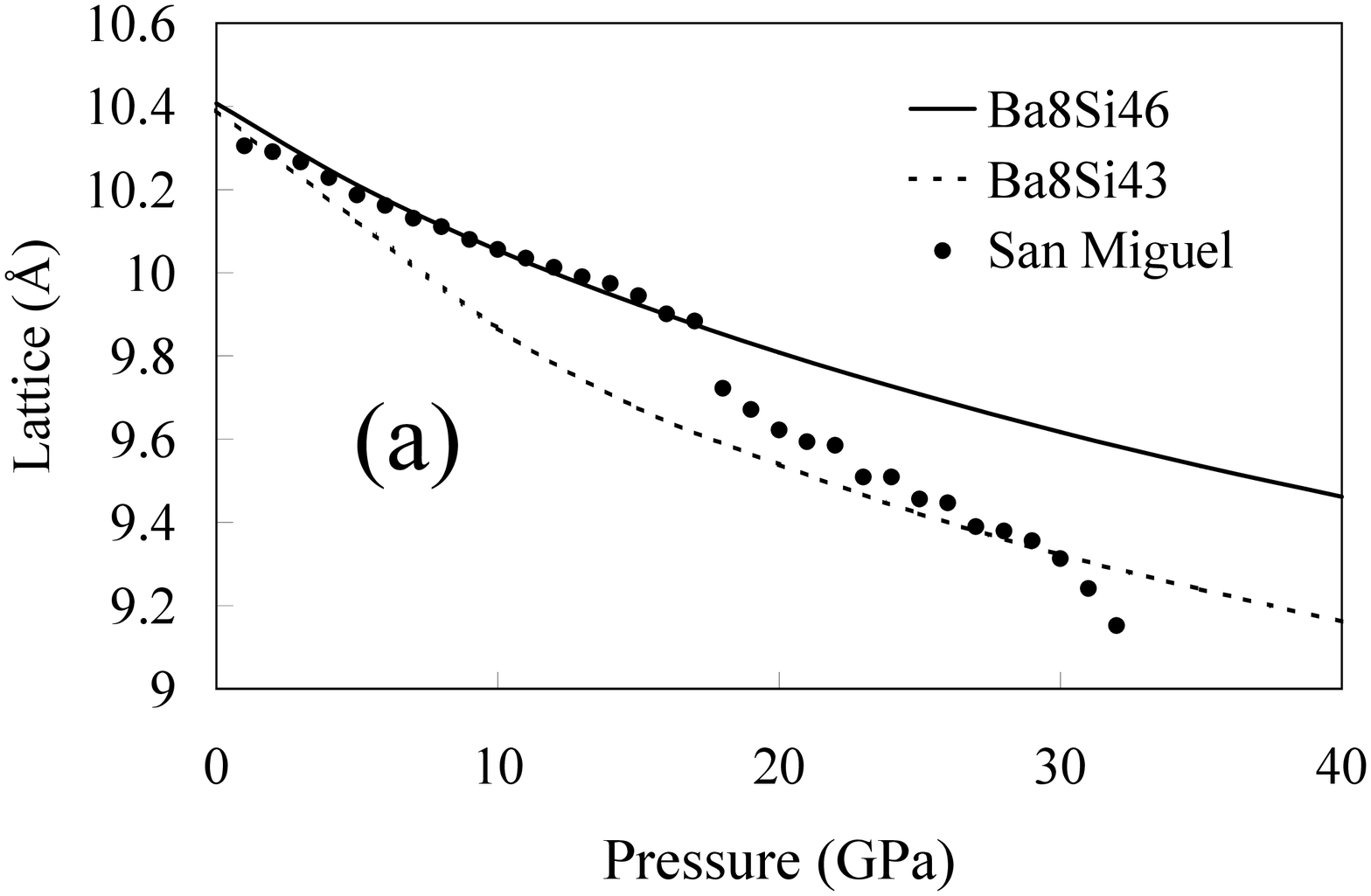}}
\resizebox{0.5\textwidth}{!}{\includegraphics*{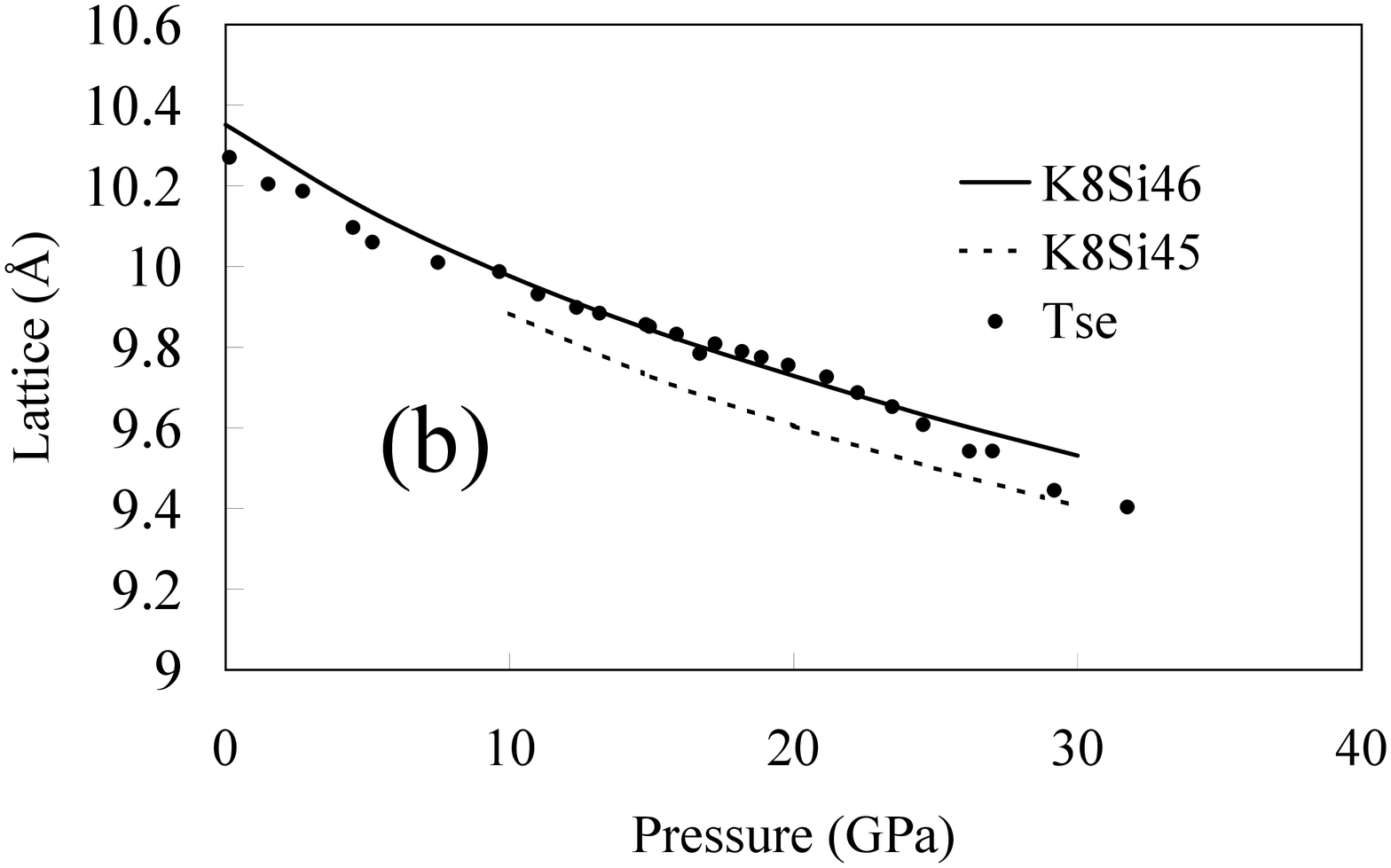}}
\end{center}
\caption{
\label{fig:EOS}
(a) Equation of state for the nominal $\mathrm{Ba_{8}Si_{46}}$.
Symbols indicate the experimental data \cite{SanMiguel2005}. Solid lines indicate the theoretical results for $\mathrm{Ba_{8}Si_{46}}$, and the dashed lines for $\mathrm{Ba_{8}Si_{43}}$. 
(b) Equation of state for the nominal $\mathrm{K_{8}Si_{46}}$. 
Symbols indicate the experimental data \cite{Tse2000}. Solid lines indicate the theoretical results for $\mathrm{K_{8}Si_{46}}$, and the dashed lines for $\mathrm{K_{8}Si_{45}}$. 
}

\end{figure}

Recently, a pressure-induced isostructural phase transition has been reported for $\mathrm{Ba_{8}Si_{46}}$ \cite{SanMiguel2002,Kume2003,SanMiguel2005} and for $\mathrm{K_{8}Si_{46}}$ \cite{ Tse2002,Kume2004}. On one hand, X-ray diffraction measurements show a sudden change of compressibility (softening) in the lattice constant around 14 GPa for $\mathrm{Ba_{8}Si_{46}}$  \cite{SanMiguel2005} and around 23 GPa for $\mathrm{K_{8}Si_{46}}$ \cite{Tse2002} (see the symbols in Fig.~\ref{fig:EOS}). Interestingly, the symmetry of the crystal is kept unchanged by this transition.  On the other hand, Raman spectrum measurements show a drastic change in vibrational states around 15 GPa for $\mathrm{Ba_{8}Si_{46}}$ (see Fig.~2 of \cite{Kume2003}) and around 20 GPa for $\mathrm{K_{8}Si_{46}}$ \cite{Kume2004}, which probably corresponds to the phase transition observed in the X-ray diffraction measurement. On the compression process of $\mathrm{Ba_{8}Si_{46}}$ \cite{Kume2003} the low energy peak about 100$\mathrm{cm^{-1}}$ and some other peaks suddenly disappears at 15 GPa resulting in six broadened Raman bands in high-pressure phase. On the decompression process the Raman spectra recover to those obtained before compression \cite{Kume2003} in spite of the drastic spectral change and the volume collapse. The disappearance and reappearance of the peak around 100 $\mathrm{cm^{-1}}$ of $\mathrm{Ba_{8}Si_{46}}$ at 15 GPa is an important signature of this phase transition. 
In addition to this higher-pressure phase transition, another phase transition has been observed in Raman spectra at a lower pressure, i.e., 7 GPa for $\mathrm{Ba_{8}Si_{46}}$ \cite{Kume2003} and 15 GPa for $\mathrm{K_{8}Si_{46}}$ \cite{Kume2004}, whose signature seems absent in the X-ray diffraction \cite{SanMiguel2005}.  The change of spectra at the lower-pressure phase transition is modest compared to the higher-pressure phase transition, and is presently interpreted as the result of minor displacements of atoms from their ideal position \cite{Kume2003,Tse2002}. Mechanism for the higher-pressure phase transition have been proposed as a change of hybridization between Ba atoms and the silicon cages by San~Miguel et al. \cite{SanMiguel2005} and as an electronic topological transition (ETT) by J.S.~Tse et al. \cite{Tse2005Ma}, however, the observed collapse of the lattice parameter has not yet been explained quantitatively by these mechanisms.


In this article, we propose a new model that explains many aspects of the observed data such as lattice constants, Raman spectra, and the effect of noble metal substitution. Our model assumes that vacancies start forming in the silicon framework, say at Si($6c$) sites, above the transition pressure ( e.g. 15 GPa for Ba). Si($6c$) sites are partially occupied by silicon atoms with random configurations, and the crystal transforms to $\mathrm{Ba_{8}Si_{46-n}}$, where $0<n<6$ and typically $n\simeq3$ at the highest pressures, i.e., 
\begin{equation}
\mathrm{Ba_{8}Si_{46}} \rightarrow \mathrm{Ba_{8}Si_{46-n}}+n \mathrm{Si}
.
\label{eq:1}
\end{equation}
 The assumption of random configuration implies that, on the average, only the occupation number of Si($6c$) sites changes to $(6-n)/6$ and the space group $Pm\overline{3}n$ is preserved. On one hand, however, the crystal may become more compressible due to the created vacancies, resulting in the collapse of the lattice parameter. On the other hand, the partial occupancy of Si($6c$) sites implies weaker binding between small cages, which may result in the softening of some Raman modes. These features will be confirmed in the following by the first principles calculations.
Actually, this type of partial occupation has been observed for the nominal $\mathrm{Ba_{8}Ge_{46}}$, an analogue material of  $\mathrm{Ba_{8}Si_{46}}$, at ambient conditions \cite{Herrmann1999}. The favorable stoichiometry is reported as $\mathrm{Ba_{8}Ge_{43}}$ \cite{Herrmann1999}.


To examine the validity of our model, we first made geometric optimization of the unit cell for $\mathrm{Ba_{8}Si_{46-{\it n}}}$, $n=0,1,3,6$ under pressure. The first-principles calculations were performed with CASTEP 4.2 codes based on the plane wave basis set\cite{Payne1992}, the Vanderbilt-type ultrasoft pseudopotentials \cite{Vanderbilt1990} for electron-ion interaction, and GGA-PBE \cite{Perdew1996} for exchange-correlation interaction. The cut off energy was set to 160 (eV) and the Brillouin zones were sampled with $2\times2\times2$ k-points. For $n=1$ ($n=3$), the partially occupied Si(6$c$) sites with random configurations were further modeled by the ordered unit cell with one (three) out of six 6$c$ sites are occupied by Si atoms with maximum symmetry. As the result, the unit cell for $n=1$ becomes tetragonal with $P\overline{4}m2$ symmetry, and  for $n=3$ trigonal with $R32$ symmetry. After geometric optimization, it was found that these unit cell is close to the original $Pm\overline{3}n$ symmetry and may approximate random occupation well. 
Symbols in Fig.~\ref{fig:EOS}a show the measured equation of state (EOS) of the nominal $\mathrm{Ba_{8}Si_{46}}$ \cite{SanMiguel2005}.
The measured EOS starts to deviate from the calculated EOS of $\mathrm{Ba_{8}Si_{46}}$ around 15 GPa and approaches to the calculated EOS of $\mathrm{Ba_{8}Si_{43}}$  above 20 GPa. This result strongly supports our scenario mentioned above. 

\begin{figure}
\begin{center}
\resizebox{0.5\textwidth}{!}{\includegraphics*{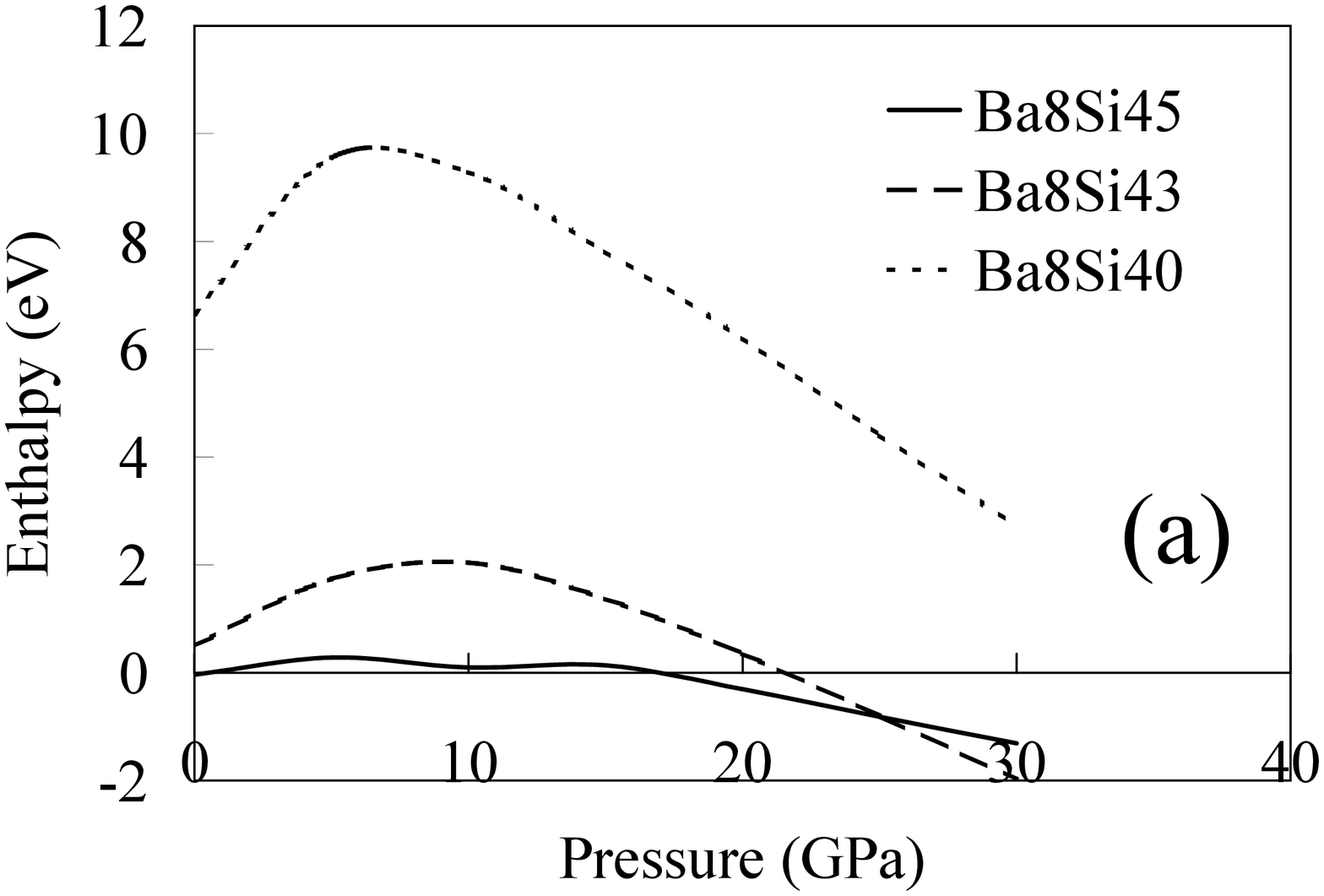}}
\resizebox{0.5\textwidth}{!}{\includegraphics*{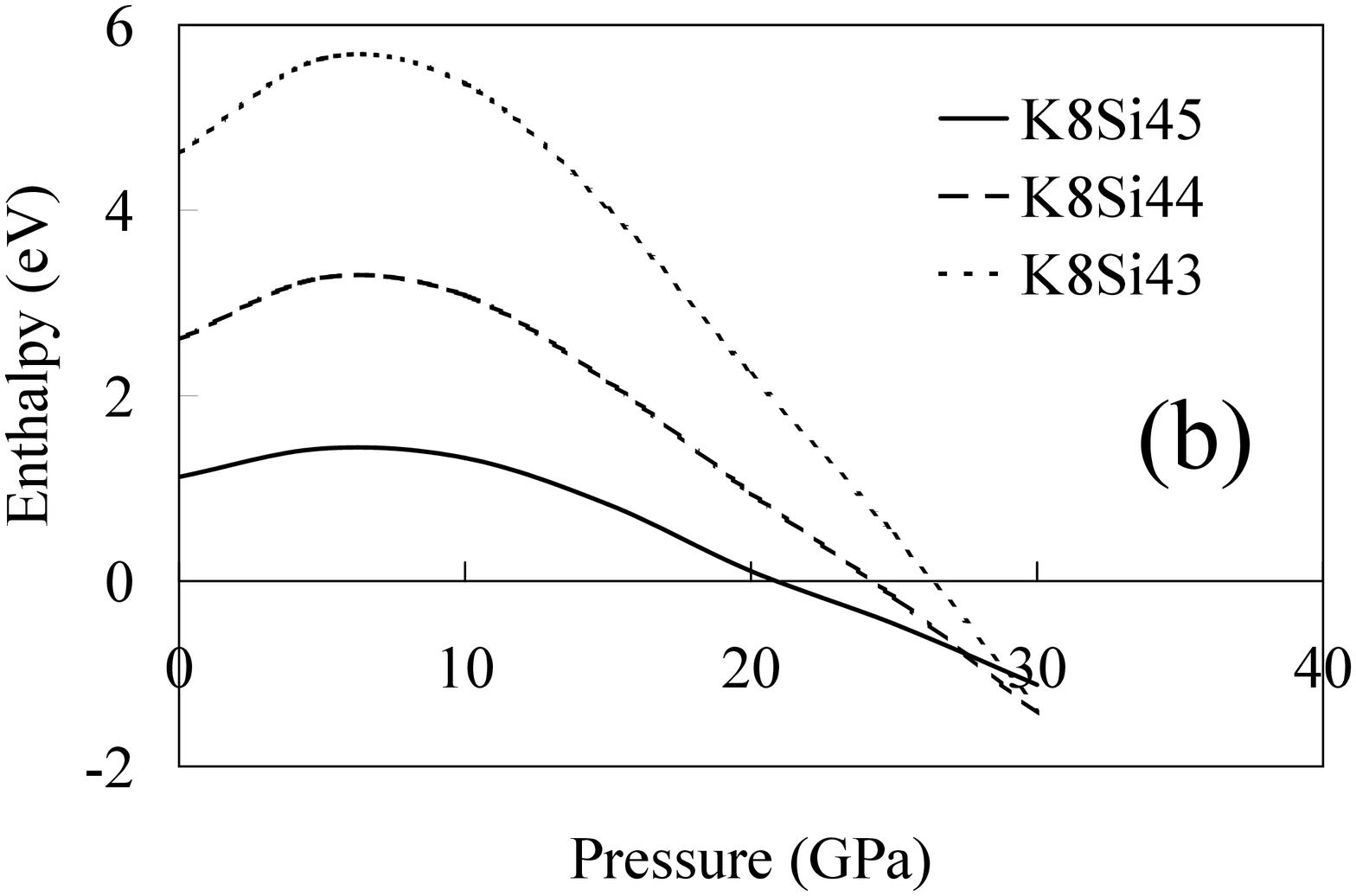}}
\end{center}
\caption{
(a) Enthalpy of $\mathrm{Ba_{8}Si_{40}}$, $\mathrm{Ba_{8}Si_{43}}$ and $\mathrm{Ba_{8}Si_{45}}$ relative to $\mathrm{Ba_{8}Si_{46}}$ estimated by first principles calculation. 
(b) Enthalpy of $\mathrm{K_{8}Si_{43}}$, $\mathrm{K_{8}Si_{44}}$ and $\mathrm{K_{8}Si_{45}}$ relative to $\mathrm{K_{8}Si_{46}}$. 
} 
\label{fig:enthalpy}
\end{figure}

Next, we study the enthalpy for vacancy formation to understand why vacancies start forming around 15 GPa and why $n$ approaches to $\simeq 3$.
For this purpose, we define the enthalpy of $\mathrm{Ba_{8}Si_{40}}$, $\mathrm{Ba_{8}Si_{43}}$ and $\mathrm{Ba_{8}Si_{45}}$ relative to $\mathrm{Ba_{8}Si_{46}}$  as follows.
\begin{equation}
\label{eq:DH}
\Delta H(\mathrm{Ba_{8}Si_{46-{\it n}}}) = H(\mathrm{Ba_{8}Si_{46-{\it n}}})+\ n H(\mathrm{Si}) -H(\mathrm{Ba_{8}Si_{46}}) 
\end{equation}
where $H(\mathrm{Ba_{8}Si_{46-{\it n}}})$ is the enthalpy of $\mathrm{Ba_{8}Si_{46-{\it n}}}$, and $H(\mathrm{Si})$ is the enthalpy per atom of the diamond silicon (below 10 GPa) and silicon V phase (above 10 GPa) \cite{Hu1984,Olijnyk1984}. For simplicity, we have ignored the intermediate minor phases between the two phases. The results are shown in Fig.~\ref{fig:enthalpy}a. $\Delta H(\mathrm{Ba_{8}Si_{45}})$ becomes negative around 15 GPa, indicating the onset of vacancy formation at Si(6$c$) sites.  Then  $\Delta H(\mathrm{Ba_{8}Si_{43}})$ becomes negative around 22 GPa, while  $\Delta H(\mathrm{Ba_{8}Si_{40}})$ remains positive even above 30 GPa. Therefore vacancies start forming around 15 GPa, and then Si(6$c$) sites approaches to the half occupation but will not be fully unoccupied. 
When the pressure changes, the number of vacancies at Si(6$c$) sites is adjusted to minimumize the enthalpy $\Delta H(\mathrm{Ba_{8}Si_{46-{\it n}}})$ .
This scenario explains the phase transition of $\mathrm{Ba_{8}Si_{46}}$ measured at 15 GPa \cite{SanMiguel2005} very well. This phase transition may be induced by the phase transition of the material composing the clathrate, i.e., Si-diamond $\rightarrow$ Si-V; Si atoms are squeezed from the clathrate because the more stable Si phase appeared. Similar phenomenon is known for the pressure-induced phase transition of methane hydrate \cite{Loveday2001,Hirai2001,Shimizu2002,Ikeda2003,Iitaka2003} where water molecules are squeezed from the hydrate crystal in accordance with the pressure-induced phase transitions of water ice. 

The electronic configuration of atomic Ba is $[\mathrm{Kr}](4d)^{10}(5s)^2(5p)^6(5d)^0(6s)^2$. However, Ba~$6s$ states are not occupied in $\mathrm{Ba_{8}Si_{46}}$: part of these two electrons occupy the hybridized states of Ba~$5d$ and Si~$3p$ states that form the narrow DOS peak at the Fermi energy, and the rest of the electrons are donated to the Si framework \cite{Kamakura2005}. Therefore Ba is {\em partially} ionized.  When one Si atom is removed from the crystal by forming a vacancy, four dangling bonds are formed on the Si framework, and these dangling bonds should be satisfied by consuming four electrons transferred from the Ba atoms\cite{Herrmann1999}. Since there are eight Ba atoms in the unit cell and each Ba atom can donate two '$6s$' electrons, the maximum number of vacancy formation per unit cell becomes $(8\times2)/4=4$. Creating more vacancies will leave unsaturated dangling bonds, resulting in much higher energy of formation. The extremely high enthalpy of $\mathrm{Ba_{8}Si_{40}}$ compared to $\mathrm{Ba_{8}Si_{45}}$ and $\mathrm{Ba_{8}Si_{43}}$ in Fig.~\ref{fig:enthalpy}a can be explained by this mechanism. Further, the increase of the width and the area of Ba $L_{III}$ X-ray absorption near-edge white line spectra as a function of pressure above 11 GPa \cite{SanMiguel2005} may be interpreted as a result of vacancies randomly generated on the Si framework. 

Now, let us discuss vacancy formation of $\mathrm{K_{8}Si_{46} }$.  
Symbols in Fig.~\ref{fig:EOS}b show the measured EOS of the nominal $\mathrm{K_{8}Si_{46}}$ \cite{Tse2000}.  The measured EOS starts to deviate from the calculated EOS of $\mathrm{K_{8}Si_{46}}$ around 20 GPa and approaches to the calculated EOS of $\mathrm{K_{8}Si_{45}}$  around 30 GPa. 
Remember that the electronic configuration of atomic K is $[\mathrm{Ar}] (4s)^1$.  Then the maximum number of vacancy formation per unit cell becomes $(8\times1)/4=2$. 
 Fig.~\ref{fig:enthalpy}b shows the enthalpy $\Delta H(\mathrm{K_{8}Si_{46-n}})$ as a function of pressure. It predicts that vacancy formation starts at 20 GPa and the crystal remains approximately $\mathrm{K_{8}Si_{45}}$ up to 30 GPa.
This result is consistent with the measured EOS. The degree of volume collapse seems controlled by valence number of the cation.



In order to explain the observed Raman spectra, we calculated the normal modes of $\mathrm{Ba_{8}Si_{46}}$ by using density functional linear response theory \cite{Baroni1987,Giannozzi1991,pwscf}. 
For each mode, we examined the symmetric characters and the contributions from each atoms. 
The mode around 100 $\mathrm{cm^{-1}}$, which is an important signature of the phase transition at 15 GPa, consists of small cages rotating around their center, which are then tied up to a collective mode being bridged by Si(6$c$) and Si(16$i$) atoms. According to our assumption, Si atoms are partially and randomly occupied above 15 GPa, and therefore the restoration force of this collective mode will become weaker and randomly distributed, resulting in a red shifted and broadened peak. 
The observed disappearance of the peak around 100 $\mathrm{cm^{-1}}$ and appearance of the broad peak around 60 $\mathrm{cm^{-1}}$ above 15 GPa may be interpreted as the result of the partial occupation of the bridging atoms. 

Ternary Si clathrate $\mathrm{Ba_{8}Au_{6}Si_{40}}$ is synthesized under high pressure from the stoichiometric mixture of $\mathrm{Ba}$, $\mathrm{Au}$ and $\mathrm{Si}$, where Si($6c$) atoms of  $\mathrm{Ba_{8}Si_{46}}$ are replaced by Au atoms  \cite{Herrmann1999}. This indicates that the Si(6$c$) atoms are weakly bound to the framework, and replacing Si(6$c$) by Au atoms makes the system more stable. Our first principles calculations of enthalpy show that $\mathrm{Ba_{8}Au_{6}Si_{40}}$ is $5\sim10$ eV more stable than $\mathrm{Ba_{8}Si_{46}}$ in this pressure range. According to our model, this stabilization of 6$c$ atoms should suppress the creation of vacancies and hence the isostructural phase transition. Actually, recent Raman spectra measurement revealed that $\mathrm{Ba_{8}Au_{6}Si_{40}}$ is stable up to 27 GPa with no phase transition observed \cite{Kume2005}.Similar stabilization of $\mathrm{Ba_{8}Ag_{6}Si_{40}}$ has been observed by X-ray diffraction \cite{SanMiguel2005a}.

So far we have discussed the energetics of the isostructural transformation.  Let us now examine the kinetics of this transformation.  According to our model the excess Si atoms are transported by vacancy mechanism. Vacancies are created at the surface or grain boundary of $\mathrm{Ba_{8}Si_{46}}$ particles by releasing the excess Si atoms from the particle and then the created vacancies propagate into the particle.
\begin{figure}
\begin{center}
\resizebox{0.5\textwidth}{!}{\includegraphics*{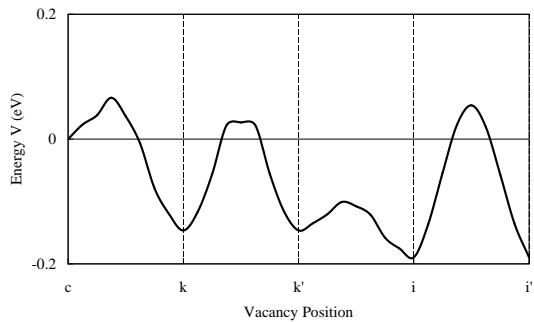}}
\end{center}
\caption{Enthalpy of a vacancy relative to the formation enthalpy at $6c$ site ( $P$=15 GPa ) is presented along the covalent framework.
\label{fig:path}
}
\end{figure}
In Fig.~\ref{fig:path}, the enthalpy of a vacancy at 15 GPa is shown as a function of position along the covalent framework, where V(6c) is set to the origin of enthalpy. In the graph, for example, the vacancy moving from the 6$c$ site to the 24$k$ site means the Si atom at 24$k$ site moves to the empty 6$c$ site.  Since equation (\ref{eq:DH}) defines $\Delta H(\mathrm{Ba_{8}Si_{45}})=V(6$c$)$, vacancies start to form at site $r$ when $\Delta H(\mathrm{Ba_{8}Si_{45}})+V(r)<0$. In contrast to $\mathrm{Ba_{8}Ge_{43}}$, it turned out that vacancies favor energetically the 16$i$ and 24$k$ sites than 6$c$ sites. When vacancies start to form at these crystallographic sites, the migration mechanism of vacancies is the thermal hopping between the sites separated by the potential barriers of c.a. 0.2 eV, which is smaller than the 0.4 eV barrier of the diamond silicon at ambient pressure \cite{El-Mellouhi} but still large enough to make the diffusion slow at ambient temperature. However, once the enthalpy difference becomes large enough, 
\begin{equation}
\label{eq:barrier}
\Delta H(\mathrm{Ba_{8}Si_{45}}) < -0.1 \mbox{(eV)},
\end{equation}
to override the barriers,  vacancies can propagate ballistic into the particle. This simple mechanism may explain the strong mobility and diffusion necessary for our model, while more realistic models may take into account the effects of energy dissipation and many vacancies. Since the condition (\ref{eq:barrier}) is almost equivalent to the vacancy formation condition for 6$c$ sites, $\Delta H(\mathrm{Ba_{8}Si_{45}}) < 0$, the discussions made with the assumption that 6$c$ site is the most favorable vacancy position, are still valid in this case. 


In summary, by using the first principles calculations, we have shown that the model of partially occupied Si sites can explain many aspects of the phase transition of $\mathrm{Ba_{8}Si_{46}}$ at 15 GPa, such as  the  equation of state, transition pressure and change of Raman spectra as well as effects of substituting Si(6$c$) atoms with noble metals.  Recently, Carrillo-Cabrera {\it et al.} \cite{Carrillo-Cabrera2004} and Okamoto {\it et al.} \cite{Okamoto2006} have reported ordered models of $\mathrm{Ba_{8}Ge_{43}}$ with $2 \times 2 \times 2$ superstructure. Our discussions of vacancy formation should be valid also for such superstructures.  For further confirming, improving, or discarding our model, more experimental and theoretical studies are necessary, especially on the reversibility of the phase transition and the fate of the released Si atoms. Precise X-ray diffraction measurement may provide a direct evidence of the partially occupied Si sites or that of the superstructures.
Raman spectrum measurement of the analogue material $\mathrm{Ba_{8}Ge_{43}}$ at ambient and high pressure conditions will provide a clue to interpret the Raman spectra of the nominal $\mathrm{Ba_{8}Si_{46}}$ above 15 GPa as those of $\mathrm{Ba_{8}Si_{43}}$ phase. 

We thank J.S.~Tse, H.~Shimizu, T.~Kume, S.~Sasaki, K.~Tanigaki, T.~Takabatake, S.~Yamanaka and T.~Ebisuzaki for useful discussions and encouragements as well as D.M.~Bird for CASTEP codes. The calculations were performed with Super-Combined Cluster system at RIKEN and supercomputers at ISSP.


\end{document}